\documentclass{article}
\usepackage{amsfonts}
\usepackage{graphicx}
\usepackage{graphics}
\newcommand{\R}{\mathbb{R}}

\begin{document}

\markboth{Fionn Murtagh}{Ultrametric and Generalized Ultrametric in 
Computational Logic and in Data Analysis}

\title{Ultrametric and Generalized Ultrametric in Logic and in Data Analysis}

\author{Fionn Murtagh \\
Science Foundation Ireland \\
Wilton Park House, Wilton Place, Dublin 2, Ireland \\
and \\
Department of Computer Science \\
Royal Holloway, University of London \\
Egham TW20 0EX, UK \\
fmurtagh@acm.org}
\maketitle

\begin{abstract}
Following a review of metric, ultrametric and generalized ultrametric, 
we review their application in data analysis.  We show how they 
allow us to explore both geometry and topology of information, starting
with measured data.  Some themes are then developed based on the 
use of metric, ultrametric and generalized ultrametric in logic.
In particular we study approximation chains in an ultrametric or 
generalized ultrametric context.  Our aim in this work is to 
extend the scope of data analysis by facilitating reasoning based on
the data analysis; and to show how quantitative and qualitative 
data analysis can be incorporated into logic programming.    
\end{abstract}

\section{Introduction}

The applicability of metric spaces to applications related to logic has 
long been known.  For example Lawvere \cite{lawvere1,lawvere2} starts 
with the observation of the analogy of the triangular inequality and 
a categorical composition law.  A comprehensive 
survey of this area can be found  in \cite{sedacj}.  

Hierarchies as used in data analysis are presented in terms of 
finding various forms of symmetry in data in \cite{murtsymm}.
We could describe hierarchy built from pairwise dissimilarities as a
``precision tool'' for data mining; and hierarchies built from
the generalized ultrametric (see section \ref{genum}) as leading to a
``power tool'' for data mining.  The former is (without special
algorithmic speedups) typically quadratic or $O(n^2)$ in its computational
requirement.  The latter can be linear or $O(n)$ in its computation.  Here
$n$ relates to number of observations.

We begin in section \ref{pairdiss} with data analysis.  We motivate
the hierarchical structuring of data, describing at a general level
how the geometry and the topology of information come into play, 
related respectively to metric and ultrametric embedding of data.  

In section \ref{sectapprox} we show how hierarchy, induced from data, 
can be made use of for approximating data.  The latter, approximating
data, is applicable and important for computational purposes.

In logic, chains of implications or conditionals have to be analyzed.
When we consider a partial order of conditionals, then the framework
of spherical (ultrametric) completeness or inductive limit
(sections \ref{sect332} and
especially \ref{dilat}) become very useful indeed.

In section \ref{sect332}, we will look at how, \cite{edalat3},
a ``{\em computable real number}
is ... the lub [least upper bound] of a shrinking sequence of rational
intervals which is generated by a master program'', and therefore how a real
number is computable ``in the interval approach 
to computability on the real line''.

The convergence to fixed points that are based on a generalized
ultrametric system is precisely the study of
spherically complete systems and expansive automorphisms
discussed in section \ref{dilat}.  As expansive automorphisms we see here
again an example of data and information symmetry at work.

\section{From Metric to Ultrametric Topology}
\label{pairdiss}

We will discuss how an ultrametric topology -- a tree structuring of
the data -- is induced from data, using pairwise dissimilarities.

\subsection{Pairwise Dissimilarities}
\label{sect621}

Given an observation set, $X$, we define dissimilarities as the mapping
$d : X \times X \longrightarrow \R^+$, where $\R^+$ are the positive reals.
A dissimilarity is a positive, definite, symmetric measure (i.e., $d(x, y)                      
\geq 0; d(x,y) = 0 \mbox{ if } x = y; d(x,y) = d(y,x)$).  If in addition
the triangular inequality is satisfied (i.e., $d(x,y) \leq d(x,z) + d(z,y),  
\forall x, y, z \in X$) then the dissimilarity is a distance.

\subsubsection{From Dissimilarities to an Ultrametric}

If $X$ is endowed with a metric, then we now describe how this metric is
mapped onto an ultrametric.  In practice, there is no need for $X$ to be
endowed with a metric.  Instead a dissimilarity is satisfactory.

A hierarchy, $H$,
is defined as a binary, rooted, node-ranked tree, also
termed a dendrogram \cite{benz,john,lerm,murt85}.
A hierarchy defines a set of embedded subsets of a given set of objects
$X$, indexed by the set $I$.
These subsets are totally ordered by an index function $\nu$, which is a
stronger condition than the partial order required by the subset relation.
A bijection exists between a hierarchy and an ultrametric space.

Let us show these equivalences between embedded subsets, hierarchy, and
binary tree, through the constructive approach of inducing $H$ on a set
$I$.

Hierarchical agglomeration on $n$ observation vectors with indices
$i \in I$ involves
a series of $1, 2, \dots , n-1$ pairwise agglomerations of
observations or clusters, with the following properties.  A hierarchy
$H = \{ q | q \in 2^I \} $ such that (i) $I \in H$, (ii) $i \in H \ \forall                     
i$, and (iii) for each $q \in H, q^\prime \in H: q \cap q^\prime \neq                           
\emptyset \Longrightarrow q \subset  q^\prime \mbox{ or }  q^\prime                             
 \subset q$.  Here we have denoted the power set of set $I$ by $2^I$.
An indexed hierarchy is the pair $(H, \nu)$ where the positive
function defined on $H$, i.e., $\nu : H \rightarrow \R^+$, satisfies:
$\nu(i) = 0$ if $i \in H$ is a singleton; and  $q \subset  q^\prime 
\Longrightarrow \nu(q) < \nu(q^\prime)$.  Here we have denoted the
positive reals, including 0, by $\R^+$.
Function $\nu$ is the agglomeration
level.  Take  $q \subset  q^\prime$, let $q \subset q''$
 and $q^\prime \subset q''$, and let $q''$ be the lowest level cluster for
which this is true. Then if we define $D(q, q^\prime) = \nu(q'')$, $D$ is
an ultrametric.  In practice, we start with a Euclidean or alternative
dissimilarity, use some criterion such as minimizing the change in variance
resulting from the agglomerations, and then define $\nu(q)$ as the
dissimilarity associated with the  agglomeration carried out.

\subsection{Metric and Ultrametric for Geometry and Topology of 
Information}

\begin{figure}
\includegraphics[width=7cm]{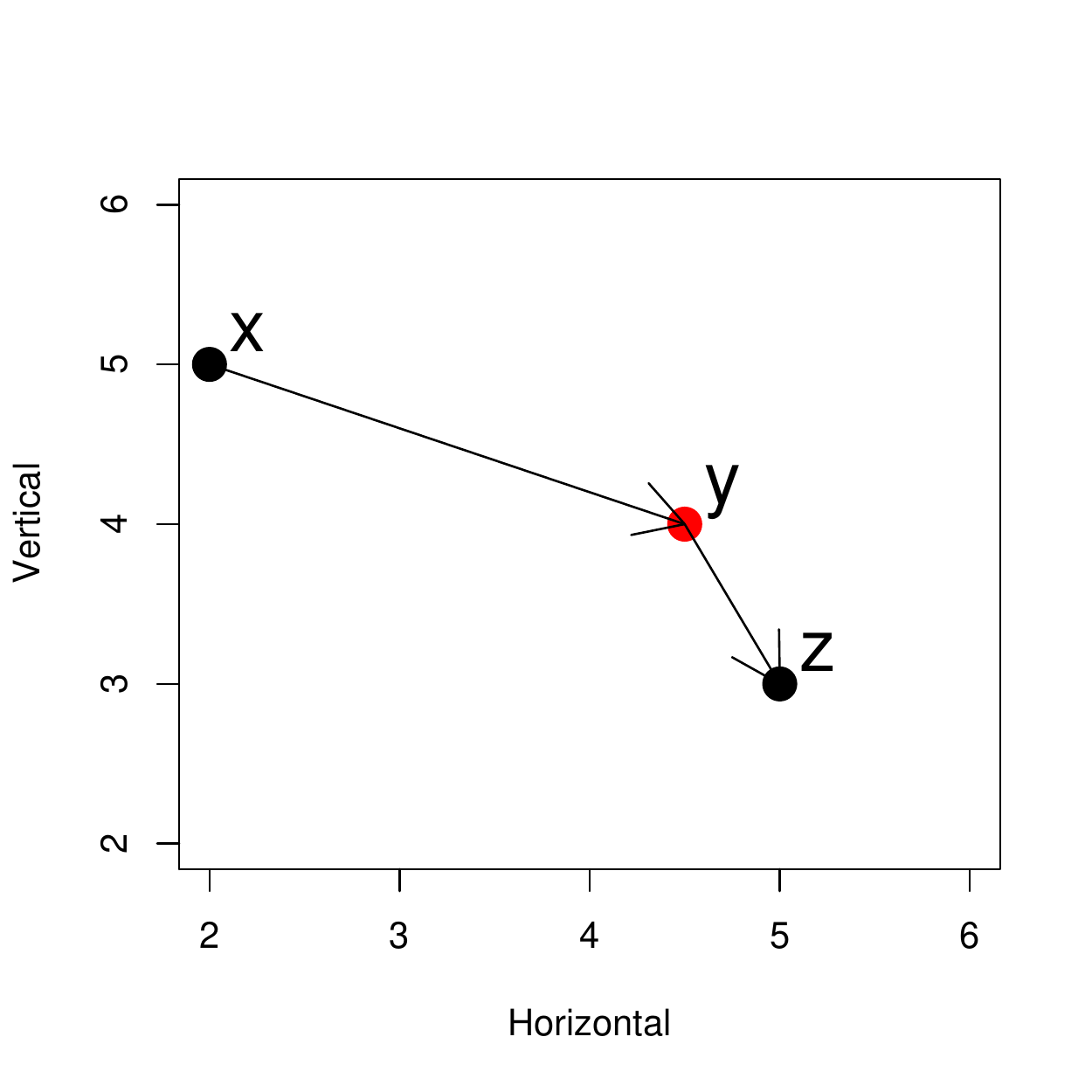}
\caption{The triangular inequality defines a metric:
every triplet of points satisfies the relationship:
$d(x,z) \leq d(x,y) + d(y,z)$ for distance $d$.}
\label{fig2}
\end{figure}

The {\em geometry of information}
is a term and viewpoint used by \cite{vanr}.
The triangular inequality holds for metrics.  An example of a metric is
the Euclidean distance, exemplified in Figure \ref{fig2}, where each
and every triplet of points satisfies the relationship:
$d(x,z) \leq d(x,y) + d(y,z)$ for distance $d$.  Two other relationships
also must hold.  These are symmetry and positive definiteness, respectively:
$d(x,y) = d(y,x)$, and $d(x,y) > 0 $ if $x \neq y$, $d(x,y) = 0 $ if $x = y$.

We come now to a different principle: that of the {\em topology of 
information}.   The particular topology used is that of hierarchy.
Euclidean embedding provides a very good starting point to look at
hierarchical relationships.
An innovation in our work is as follows: the hierarchy takes sequence, e.g.\
timeline, into account.
This captures, in a more easily understood way, the notions of
novelty, anomaly or change.

\begin{figure}
\includegraphics[width=7cm]{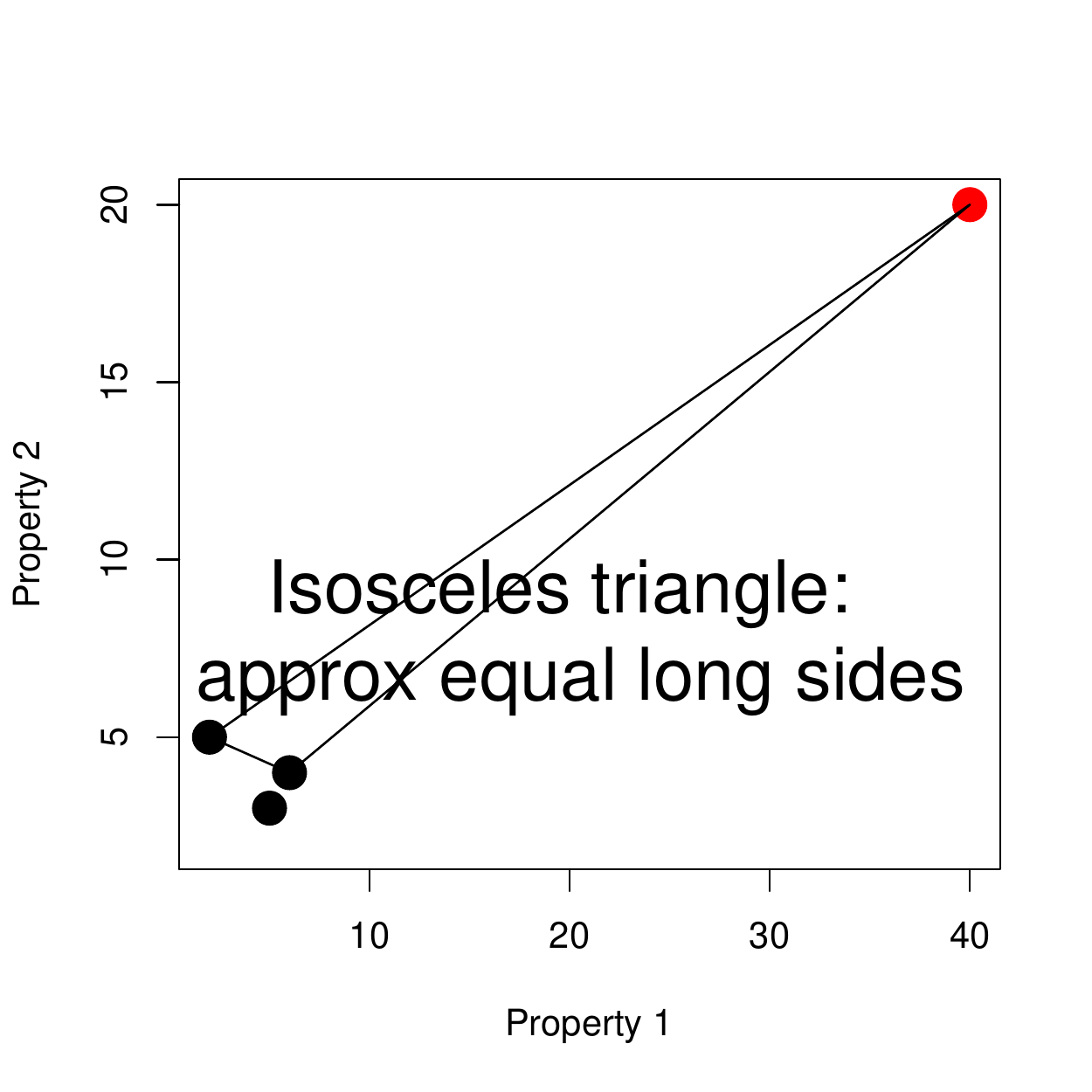}
\caption{The query is on the far right.  While we can easily
determine the closest target (among the three objects represented
by the dots on the left), is the closest really that much different
from the alternatives?}
\label{fig4}
\end{figure}

Let us take an informal case study to see how this works.
Consider the situation of seeking documents based on titles.  If the target
population has at least one document that is close to the query,
then this is (let us assume) clearcut.  However if all documents in the
target population are very unlike the query, does it make any sense to
choose the closest?  Whatever the answer here we are focusing on the
inherent ambiguity, which we will note or record in an appropriate way.
Figure \ref{fig4} illustrates this situation, where
the query is the point to the right.

By using approximate similarity this situation can be
modeled as an isosceles triangle with small base, as illustrated in
Figure \ref{fig4}.
An ultrametric space has properties that are very unlike a metric
space, and one such property is that the only triangles allowed are either
(i) equilateral, or (ii) isosceles with small base.  So Figure \ref{fig4}
can be taken as representing a case of ultrametricity.  What this means
is that the query can be viewed as having a particular sort of dominance
or hierarchical relationship vis-\`a-vis any pair of target documents.
Hence any triplet of points here, one of which is the query (defining the
apex of the isosceles, with small base, triangle), defines local hierarchical
or ultrametric structure.  (See \cite{murt04} for case studies.)

It is clear from Figure \ref{fig4} that we should use approximate
equality of the long sides of the triangle.  The further away the query is
from the other data then the better is this approximation \cite{murt04}.

What sort of explanation does this provide for our conundrum?  It means that
the query is a novel, or anomalous, or unusual ``document''.   It is up to
us to decide how to treat such new, innovative cases.  It raises though the
interesting perspective that here we have a way to model and
subsequently handle the semantics of anomaly or innocuousness.

\begin{figure}
\includegraphics[width=8cm]{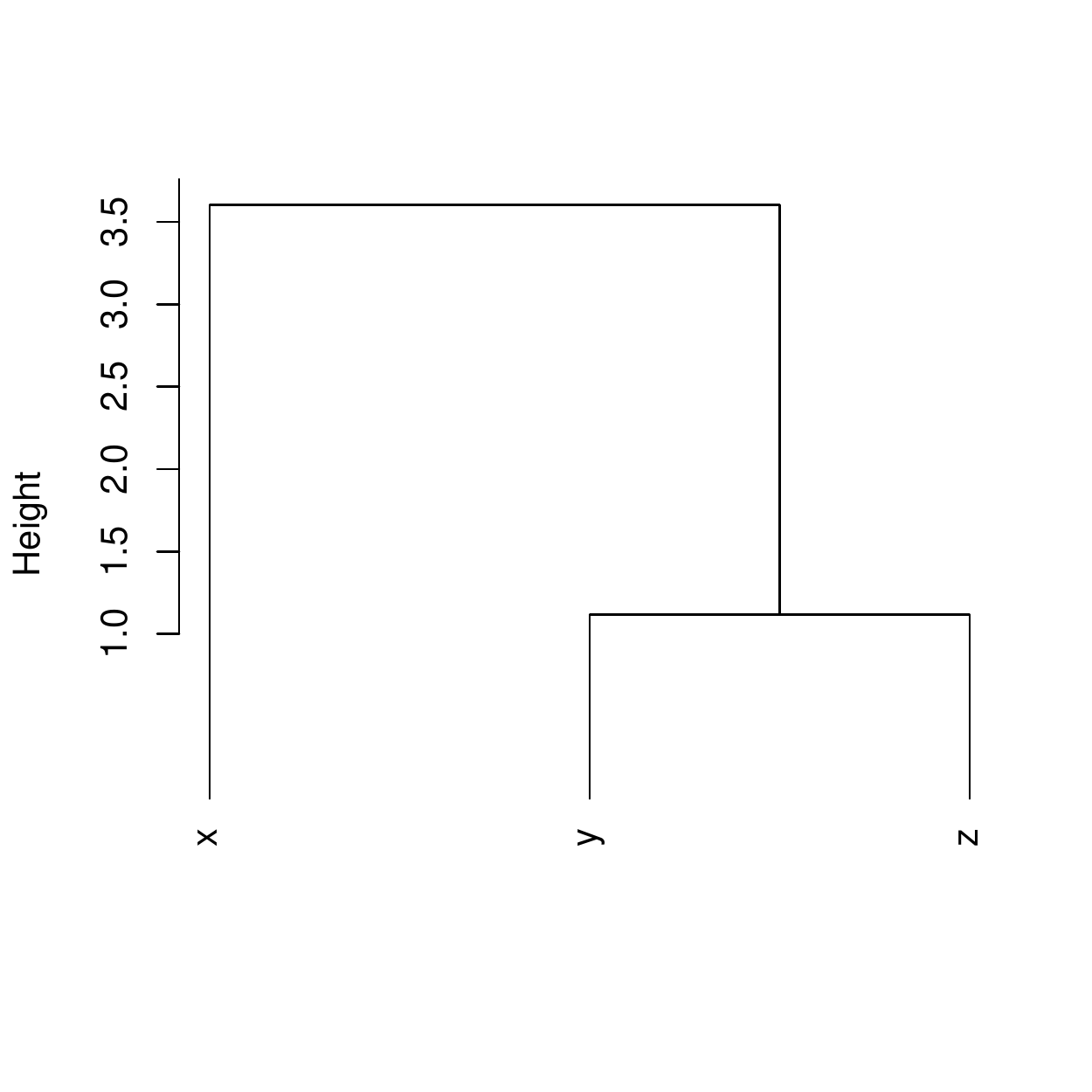}
\caption{The strong triangular inequality defines an ultrametric:
every triplet of points satisfies the relationship:
$d(x,z) \leq \mbox{max} \{ d(x,y), d(y,z) \}$ for distance $d$.
Cf.\ by reading off the hierarchy, how this is verified for all $x, y, 
z$: $d(x,z) = 3.5; d(x,y) = 3.5; d(y,z) = 1.0$.
In addition the symmetry and positive definiteness conditions
hold for any pair of points.}
\label{fig5}
\end{figure}

The strong triangular inequality, or ultrametric inequality,
holds for tree distances: see Figure \ref{fig5}.
The closest common ancestor distance is such an ultrametric.

\subsection{Hierarchical Agglomerative Clustering}

Since pairwise dissimilarities are used in constructing the hierarchy,
the computation complexity of hierarchical clustering is 
at least $O(n^2)$.
As the closest clusters (including singletons) are agglomerated at
each of $n-1$ agglomerations (card $X = $ card $I = n$),
the newly created cluster must be related to others.  This is part and parcel
of the agglomeration criterion, and can be viewed either as the cluster
update rule, or the agglomerative criterion (e.g., based on compactness,
or connectivity).

The most efficient algorithms are based on nearest neighbor chains, which
by definition end in a pair of agglomerable reciprocal nearest neighbors.
$O(n^2)$ computation time is guaranteed.  
The uniqueness and acceptability of on-the-fly agglomeration based on
reciprocal nearest neighbors can be proven (respectively, disproven) for the
given agglomerative criterion.  The reciprocal nearest neighbor algorithm
was first proposed in two articles in the journal {\em Les Cahiers de
l'Analyse des Donn\'ees} in 1980 and 1982, and are now used in software
packages such as Clustan and R.  Further information can be found in
\cite{murt83,murt84,murt85,murt92}.

\subsection{Hierarchy as the Wreath Product Group expressing Symmetries}
\label{sect4}

A dendrogram like that shown in Figure \ref{fig2} is invariant relative
to rotation (alternatively, here: permutation) of left and right
child nodes.  These rotation (or permutation) symmetries are defined
by the wreath product group (see \cite{foote1,foote2,foote3} for
an introduction and applications in signal and image processing),
and can be used with any m-ary tree,
although we will treat the binary case here.


For the group actions, with respect to which we will seek
invariance, we consider independent cyclic shifts of the
subnodes of a given node (hence, at each level).  Equivalently
these actions are adjacency preserving permutations of
subnodes of a given node (i.e., for given $q$,
with $q = q' \cup q''$, the permutations
of $\{  q', q'' \}$).
We have therefore cyclic
group actions at each node, where the cyclic group is of order
2.

The symmetries of $H$ are given by structured permutations of
the terminals.  The terminals will be denoted here by
Term $H$. The full group of symmetries is summarized
by the following generative algorithm:

\begin{enumerate}
\item For level  $l = n - 1$ down to 1 do:
\item Selected node, $\nu \longleftarrow $ node at level $l$.
\item And permute subnodes of $\nu$.
\end{enumerate}

Subnode $\nu$ is the root of subtree $H_\nu$.  We denote $H_{n-1}$ simply by
$H$.  For a subnode $\nu'$ undergoing a relocation action in step 3, the
internal structure of subtree $H_{\nu'}$ is not altered.

The algorithm described defines the automorphism group which is a
wreath product of the symmetric group.  Denote the permutation at level
$\nu$ by $P_\nu$.  Then the automorphism group is given by:
$$G = P_{n-1} \ \mathrm{wr} \ P_{n-2} \ \mathrm{wr} \ \dots \ \mathrm{wr} \ P_2
\  \mathrm{wr} \  P_1$$
where wr denotes the wreath product.

Call Term $H_\nu$ the terminals that descend from the node at level $\nu$.
So these are the terminals of the subtree $H_\nu$ with its root node at
level $\nu$.  We can alternatively call Term $H_\nu$ the cluster associated
with level $\nu$.

We will now look at shift invariance under the group action.  This amounts
to the requirement for a constant function defined on Term $H_\nu, \forall 
\nu$.  A convenient way to do this is to define such a function on the set
Term $H_\nu$ via the root node alone, $\nu$.  By definition then we have a
constant function on the set Term $H_\nu$.

Let us call $V_\nu$ a space of functions that are constant on Term $H_\nu$.
Possible bases of $V_\nu$ that were considered in \cite{murhaar} are:

\begin{enumerate}
\item Basis  vector with $| \mathrm{Term} H_{n-1} |$
components, with 0 values
except for value 1 for component $i$.

\item Set (of cardinality $n = | \mathrm{Term} H_{n-1} |$) of $m$-dimensional
observation vectors.
\end{enumerate}

The constant function for each node or level $\nu$ is:
$$ L : \mathrm{Term} H_\nu  \longrightarrow V_\nu$$

Consider the resolution scheme arising from moving from \\
$\{ \mathrm{Term} H_{\nu'}, \mathrm{Term} H_{\nu''} \}$ to
$ \mathrm{Term} H_\nu $.  From the hierarchical clustering point of view it is
clear what this represents, simply, an agglomeration of two clusters
called Term $H_{\nu'}$ and Term $H_{\nu''}$, replacing them with a new
cluster, Term $H_\nu$.

Let the spaces of constant functions
corresponding to the two cluster agglomerands be denoted $V_{\nu'}$ and
$V_{\nu''}$.  These two clusters are disjoint initially, which motivates
us taking the two spaces as a couple: $(V_{\nu'}, V_{\nu''})$.
In the same way, let the space of constant functions
corresponding to node $\nu$ be denoted $V_\nu$.

Let us exemplify a case that satisfies all that has been
defined in the context of the wreath product invariance that we are
targeting.  It is the  algorithm discussed in depth in
\cite{murhaar}.
Take the constant function on $V_{\nu'}$ to be $f_{\nu'}$.
Take the constant function on $V_{\nu''}$ to be $f_{\nu''}$.
Then define the constant function, the {\em scaling function},
 on $V_{\nu}$ to be
$(f_{\nu'} + f_{\nu''})/2$.  Next define the zero mean function,
$(w_{\nu'} + w_{\nu''})/2 = 0$, the {\em wavelet function}, as follows:

$$w_{\nu'} = (f_{\nu'} + f_{\nu''})/2 - f_{\nu'}$$
in the support interval of $V_{\nu'}$, i.e.\ Term $H_{\nu'}$, and
$$w_{\nu''} = (f_{\nu'} + f_{\nu''})/2 - f_{\nu''}$$
in the support interval of $V_{\nu''}$, i.e.\ Term $H_{\nu''}$.

Since $w_{\nu'} = - w_{\nu''}$ we have the zero mean requirement.

\section{Approximation in an Ultrametric Topology}
\label{sectapprox}

We now seek to use a hierarchical clustering for successively 
approximating an object.  In \cite{murtult} we have examples of 
application to facial recognition and textual analysis.  

Following a general view of hierarchical approximation in subsection
\ref{dilat}, we then proceed to an algorithm, and a data analysis
framework, to support hierarchical approximation.  

\subsection{Approximation from a Hierarchy: Dilation Operation as 
p-Adic Multiplication by $1/p$}
\label{dilat}

Scale-related symmetry is very important in practice.  In this
subsection we introduce an operator that provides this symmetry.  We also
term it a dilation operator, because of its role in the wavelet transform
on trees (see \cite{murhaar} for discussion and examples).

\begin{figure}
\centering
\includegraphics[width=14cm]{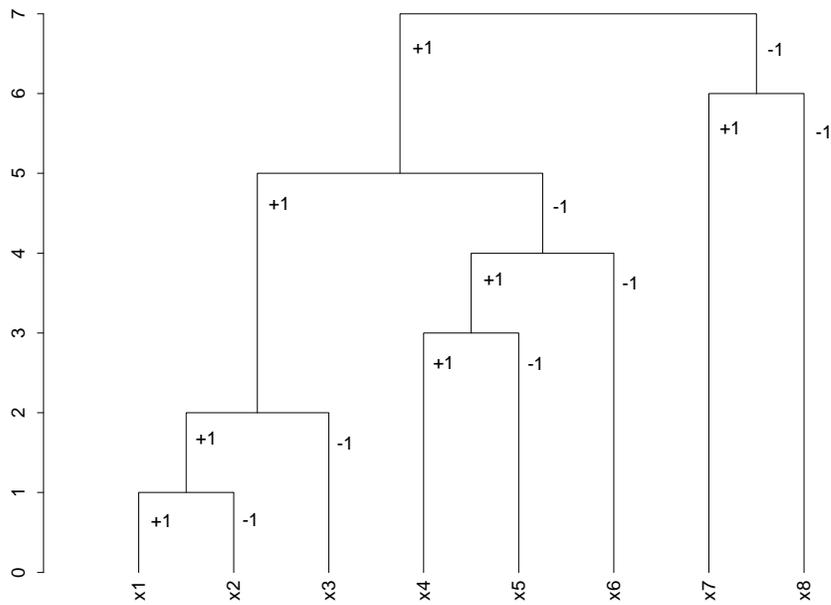}
\caption{Labeled, ranked dendrogram on 8 terminal nodes,
$x_1, x_2, \dots , x_8$.  Branches are labeled
$+1$ and $-1$. Clusters are: $q_1 = \{ x_1, x_2 \}, 
q_2 = \{ x_1, x_2, x_3 \}, q_3 = \{ x_4, x_5 \}, q_4 =  \{ x_4, x_5, 
x_6 \}, q_5 = \{ x_1, x_2, x_3, x_4, x_5, x_6 \}, 
q_6 = \{ x_7, x_8 \}, q_7 = \{ x_1, x_2, \dots , x_7, x_8 \}$.}
\label{fig222}
\end{figure}

First we introduce a p-adic encoding of a hierarchy, using Figure 
\ref{fig222} as an example.
By means of terminal-to-root traversals, we define the following
p-adic encoding of terminal nodes, and hence objects, in Figure \ref{fig222}.

\begin{eqnarray}
x_1: & + 1 \cdot p^1 + 1 \cdot p^2 + 1 \cdot p^5 + 1 \cdot p^7  \\ \nonumber
x_2: & - 1 \cdot p^1 + 1 \cdot p^2 + 1 \cdot p^5 + 1 \cdot p^7  \\ \nonumber
x_3: & - 1 \cdot p^2 + 1 \cdot p^5 + 1 \cdot p^7  \\ \nonumber
x_4: & + 1 \cdot p^3 + 1 \cdot p^4 - 1 \cdot p^5 + 1 \cdot p^7   \\ \nonumber
x_5: & - 1 \cdot p^3 + 1 \cdot p^4 - 1 \cdot p^5 + 1 \cdot p^7   \\ \nonumber
x_6: & - 1 \cdot p^4 - 1 \cdot p^5 + 1 \cdot p^7  \\ \nonumber
x_7: & + 1 \cdot p^6 - 1 \cdot p^7   \\ \nonumber
x_8: & - 1 \cdot p^6 - 1 \cdot p^7
\label{eqn00}
\end{eqnarray}

If we choose $p = 2$ the resulting decimal equivalents could be the same:
cf.\ contributions based on $+1 \cdot p^1$ and $-1 \cdot p^1 + 1 \cdot p^2$.
Given that the coefficients of the $p^j$
terms ($1 \leq j \leq 7$) are in the set $\{ -1, 0, +1 \}$
(implying for $x_1$ the additional terms: $+ 0 \cdot p^3 + 0 \cdot p^4 
+ 0 \cdot p^6$),
the coding based on $p = 3$ is required to avoid ambiguity among
decimal equivalents.

Consider the set of objects
$\{ x_i | i \in I \}$ with its p-adic coding considered
above.  Take $ p = 2$.  (Non-uniqueness of corresponding decimal codes is
not of concern to us now, and taking this value for $p$ is without any
loss of generality.)
Multiplication of
$x_1 = + 1 \cdot 2^1 + 1 \cdot 2^2 + 1 \cdot 2^5 + 1 \cdot 2^7 $ by
$1/p = 1/2$ gives: $  + 1 \cdot 2^1 + 1 \cdot 2^4 + 1 \cdot 2^6$.  Each
level has decreased by one, and the lowest level has been lost.
Subject to the lowest level of the tree being lost, the form of the tree
remains the same.  By carrying
out the multiplication-by-$1/p$ operation on all objects, it is seen that
the effect is to rise in the hierarchy by one level.

Let us call product with $1/p$ the operator $A$.  The effect of losing the
bottom level of the dendrogram means that either (i) each cluster (possibly
singleton) remains the same; or (ii) two clusters are merged.  Therefore
the application of $A$ to all $q$ implies a subset relationship between
the set of clusters $\{ q \}$ and the result of applying $A$, $\{ A q \}$.

Repeated application of the operator $A$ gives $A q$, $A^2 q$,
$A^3 q$, $\dots$.  Starting with any singleton, $i \in I$, this gives
a path from the terminal to the root node in the tree.  Each such
path ends with the null element, and therefore
the intersection of the paths equals the
null element.   

Benedetto and Benedetto \cite{benben,ben}  discuss $A$ as an expansive
automorphism of $I$, i.e.\ form-preserving, and locally expansive.
Some implications \cite{benben} of  
the 
expansive automorphism follow.
For any $q$, let us
take $q, A q, A^2 q, \dots$ as a sequence of open subgroups of
$I$, with $q \subset A q \subset A^2 q \subset \dots$, and $I = 
\bigcup \{ q, A q, A^2 q, \dots \} $.  This is termed an inductive sequence
of $I$, and $I$ itself is the inductive limit
(\cite{reiter}, p.\ 131).  

Each path defined by application of the expansive automorphism
defines a spherically complete system \cite{schi,gajic,vanRooij},
which is a
formalization of well-defined subset embeddedness.

\subsection{Haar Wavelet Transform of a Dendrogram}

Determining successive approximations of data, based on the data 
itself, leads us to the Haar wavelet transform of a hierarchy, or on a
dendrogram. 

The discrete wavelet transform is a decomposition of data into spatial and
frequency components. In terms of a dendrogram these components are with
respect to, respectively, within and between clusters of successive
partitions. We show how this works taking the data of Table \ref{table5}.

\begin{table}[tbp]
\begin{center}
\begin{tabular}{|rrrrr|}
\hline
& Sepal.L & Sepal.W & Petal.L & Petal.W \\ \hline
1 & 5.1 & 3.5 & 1.4 & 0.2 \\
2 & 4.9 & 3.0 & 1.4 & 0.2 \\
3 & 4.7 & 3.2 & 1.3 & 0.2 \\
4 & 4.6 & 3.1 & 1.5 & 0.2 \\
5 & 5.0 & 3.6 & 1.4 & 0.2 \\
6 & 5.4 & 3.9 & 1.7 & 0.4 \\
7 & 4.6 & 3.4 & 1.4 & 0.3 \\
8 & 5.0 & 3.4 & 1.5 & 0.2 \\ \hline
\end{tabular}%
\end{center}
\caption{First 8 observations of Fisher's iris data. L and W refer to length
and width.}
\label{table5}
\end{table}

\begin{figure*}[tbp]
\begin{center}
\includegraphics[width=14cm]{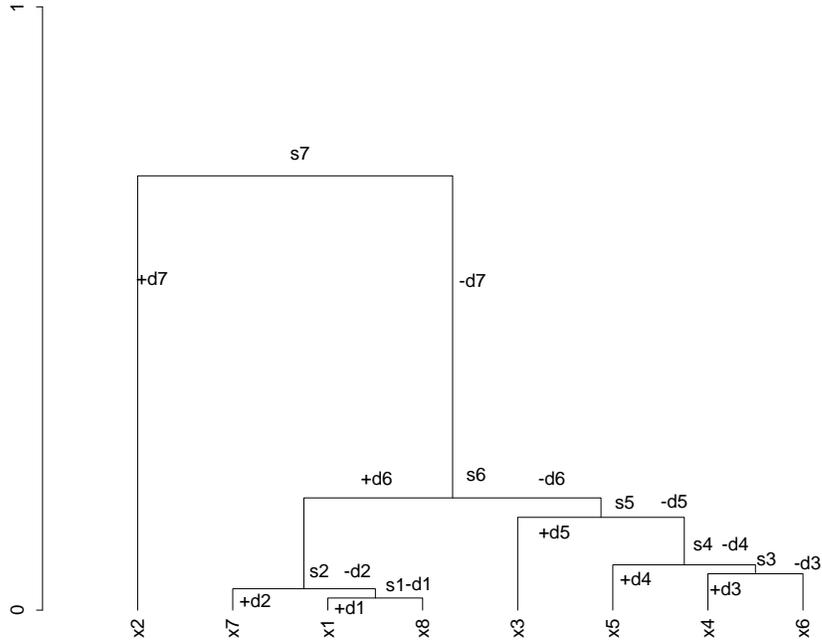}
\end{center}
\caption{Dendrogram on 8 terminal nodes constructed from first 8 values of
Fisher iris data. (Median agglomerative method used in this case.) Detail or
wavelet coefficients are denoted by $d$, and data smooths are denoted by $s$%
. The observation vectors are denoted by $x$ and are associated with the
terminal nodes. Each \emph{signal smooth}, $s$, is a vector. The (positive
or negative) \emph{detail signals}, $d$, are also vectors. All these vectors
are of the same dimensionality.}
\label{fig555}
\end{figure*}

The hierarchy built on the 8 observations of Table \ref{table5} is shown in
Figure \ref{fig555}.

Something more is shown in Figure \ref{fig555}, namely the detail signals
(denoted $\pm d$) and overall smooth (denoted $s$), which are determined in
carrying out the wavelet transform, the so-called forward transform.

The inverse transform is then determined from Figure \ref{fig555} in the
following way. Consider the observation vector $x_2$. Then this vector is
reconstructed exactly by reading the tree from the root: $s_7 + d_7 = x_2$.
Similarly a path from root to terminal is used to reconstruct any other
observation. If $x_2$ is a vector of dimensionality $m$, then so also are $%
s_7$ and $d_7$, as well as all other detail signals.

\begin{table*}[tbp]
\begin{center}
\begin{tabular}{|rrrrrrrrr|}
\hline
& s7 & d7 & d6 & d5 & d4 & d3 & d2 & d1 \\ \hline
Sepal.L & 5.146875 & 0.253125 & 0.13125 & 0.1375 & $-0.025$ & 0.05 & $-0.025$
& 0.05 \\
Sepal.W & 3.603125 & 0.296875 & 0.16875 & $-0.1375$ & 0.125 & 0.05 & $-0.075$
& $-0.05$ \\
Petal.L & 1.562500 & 0.137500 & 0.02500 & 0.0000 & 0.000 & $-0.10$ & 0.050 &
0.00 \\
Petal.W & 0.306250 & 0.093750 & $-0.01250$ & $-0.0250$ & 0.050 & 0.00 & 0.000
& 0.00 \\ \hline
\end{tabular}%
\end{center}
\caption{The hierarchical Haar wavelet transform resulting from use of the
first 8 observations of Fisher's iris data as shown in Table 
\ref{table5}. Wavelet coefficient levels are denoted d1 through d7, and the
continuum or smooth component is denoted s7.}
\label{table6}
\end{table*}

This procedure is the same as the Haar wavelet transform, only applied to
the dendrogram and using the input data.

The data required to define this wavelet transform, for the data in Table 
\ref{table5}, is shown in Table \ref{table6}.

The principle of ``folding'' the hierarchy onto an external signal is as
follows. The wavelet transform codifies the hierarchy. Having that, we apply
the ``codification'' of the hierarchy with the new, external signal as input.

Wavelet regression entails setting small and hence unimportant detail
coefficients to 0 before applying the inverse wavelet transform.

More discussion can be found in \cite{murhaar}.

\subsection{Representation of an Object as a Chain of Successively Finer
Approximations}
\label{sect3}

From the wavelet transformed hierarchy  
we can read off that, say, $x_1 = d_2 + d_5 + d_7 + s_7$:
cf.\ Figure \ref{fig555}.   Or $x_8 = d_6 - d_7 + s_7$.
These relationships use the appropriate vectors 
shown in Table \ref{table6}.  
Such relationships furnish the definitions used by
 the inverse wavelet transform, i.e.\ the recreation of the input data
from the transformed data.

Thus, the Haar dendrogram wavelet transform gives us 
 an additive decomposition of a given observation
(say, $x_1$) in terms of a degrading approximation, with a variable number
of terms in the decomposition.  The objects, or observations, are
those things which we are analyzing and on which we have (i) induced a 
hierarchical clustering, and (ii) further processed the hierarchical 
clustering in such a way that we can derive the Haar decomposition.  
In this section we will look at how this allows us to consider each object
as a limit point. 
Our interest lies in our object set, characterized by a set of data, 
 as a set of limit or fixed points.  


Using notation from domain theory (see, e.g., \cite{edalat3}) we write:

\begin{equation}
 s_7 \sqsubseteq s_7 + d_7 \sqsubseteq s_7 + d_7 + d_5
\sqsubseteq s_7 + d_7 + d_5 + d_2
\label{eqn5}
\end{equation}

The relation $a \sqsubseteq b$ is read: $a$ is an approximation to $b$,
or $b$ gives more information than $a$.  (Edalat \cite{edalat03} 
discusses examples.)
Just rewriting the very last, or rightmost, term in relation (\ref{eqn5})
gives:

\begin{equation}
s_7 \sqsubseteq s_7 + d_7 \sqsubseteq s_7 + d_7 + d_5
\sqsubseteq x_1
\label{eqn5b}
\end{equation}

Every one of our observation vectors (here, e.g., $x_1$) can be
increasingly well approximated by a {\em chain} of the sort shown in
relations (\ref{eqn5}) or (\ref{eqn5b}), 
starting with a least element ($s_7$; more generally,
for $n$ observation vectors, $s_{n-1}$).  The observation vector itself (e.g.,
$x_1$) is a least upper bound (lub) or supremum (sup), denoted
$\sqcup$ in domain theory, of this chain.  Since every observation vector
has an associated chain, 
every chain
has a lub.  The elements of 
the ``rolled down''
tree, $s_7$, $s_7 + d_7$ and $s_7 - d_7$,  $s_7 + d_7 +d_5$ and
$s_7 + d_7 - d_5$, and so on, are clearly representable as a binary rooted
tree, and the elements themselves comprise a partially ordered set (or
poset).  A {\em complete partial order} or {\em cpo} or {\em domain} is a
poset with least element, and such that every chain has a lub.  Cpos generalize
complete lattices: see \cite{davey} for lattices, domains, and their 
use in fixpoint applications.  


\subsection{Approximation Chain using a Hierarchy}

An alternative, although closely related, structure with which domains
are endowed is that of spherically complete ultrametric spaces.   The
motivation comes from logic programming, where non-monotonicity may well
be relevant (this arises, for example, with the negation operator).  Trees
can easily represent positive and negative assertions.  The general notion
of convergence, now, is related to {\em spherical completeness} 
(\cite{schi,seda}; see also \cite{khrenn1}, Theorem 4.1).  
If we have any set of embedded clusters,
or any chain, $q_k$, then the condition that such a chain be non-empty,
$\bigcap_k q_k \neq \emptyset$, means that this ultrametric space is
non-empty.  This gives us both a concept of completeness, and also a
fixed point which is associated with the ``best approximation'' of the
chain.

Consider our space of observations, $X = \{ x_i | i \in I \}$.  The
hierarchy, $H$, or binary rooted tree, defines an ultrametric space.  For
each observation $x_i$, by considering the chain from root cluster to
the observation, we see that $H$ is a spherically complete ultrametric
space.

\subsection{Mapping of Spherically Complete Space into Dendrogram Wavelet
Transform Space}

Consider analysis of the set of observations, $\{ x_i \in X \subset 
\R^m \}$.  Through use of any hierarchical clustering (subject to 
being binary, a sufficient condition for which is that a pairwise 
agglomerative algorithm was used to construct the hierarchy), followed by 
the Haar wavelet transform of the dendrogram, we have an approximation 
chain for each $x_i \in X$.  This approximation chain is defined in terms of
embedded sets.  Let $n = \mbox{card }  X$, the cardinality of the set $X$.  
Our Haar dendrogram wavelet transform
allows us to associate the set $\{ \nu_j | 1 \leq j \leq n-1 \} \subset 
\R^m$ with the chains, as seen in section \ref{sect3}.  

We have two associated vantage points on the generation of observation
$i, \forall i$: 
the set of embedded sets in the approximation chain starting always with the 
entire observation set, $I$, and ending with the singleton observation; or
the global smooth in the Haar transform, that we will call $\nu_{n-1}$, 
running through all details 
$\nu_j$ on the path, such that an additive combination of 
path members increasingly approximates the vector $x_i$ that corresponds to 
observation $i$.  Our two associated views are, respectively, a set of 
sets; or a set of vectors in $\R^m$.  We recall that $m$ is the 
dimensionality of the embedding space of our observations.  Our two 
associated views of the (re)generation of an observation  both 
rest on the hierarchical or tree structuring of our data.

\section{Generalized Ultrametric}
\label{genum}

\subsection{Applications of Generalized Ultrametrics}
\label{sect332}

As noted in the previous subsection, the usual
ultrametric is an ultrametric distance, i.e.\ for a set I,
$d: I \times I \longrightarrow \R$
(so the ultrametric distance is a real value).  The generalized
ultrametric is:
$d: I \times I \longrightarrow \Gamma$,
where $\Gamma$ is a partially ordered set.  In other words, the
{\em generalized} ultrametric distance is a set.  With this set one
can have a value,
so the usual and the generalized ultrametrics can amount to more or
less the same in practice (by ignoring the set and concentrating on its
associated value).  After all, in a dendrogram one does have
a set associated with each ultrametric distance value (and this is most
conveniently the terminals dominated by a given node; but we could have other
designs, like some representative subset or other,
of these terminals).  Remember that the set,
$\Gamma$, is defined from the original attributes (which we denote by the
set $J$); whereas the sets of observations read off a dendrogram are
subsets of the observation set (which we label with the index set $I$).
So $\Gamma = 2^J$ (and not $2^I$).

In the theory of reasoning, a  monotonic operator is rigorous application
of a succession of  conditionals (sometimes called consequence
relations).  However:
``In order to deal with programs of a more general kind (the 
so-called disjunctive programs) it became necessary to consider
multi-valued mappings'', supporting non-monotonic reasoning in the
 way now to be described (\cite{Priess0}, pp.\ 10, 13).
The novelty in the work of \cite{Priess0,Priess} is that these
authors use the
generalized
ultrametric as  a multivalued mapping.

(A more critical view of the usefulness of the generalized ultrametric
perspective is presented by \cite{kroetzsch}.)


The generalized ultrametric approach has been motived \cite{seda0}
as follows.  ``Situations arise ... in computational logic in the presence
of negations which force non-monotonicity of the operators involved''.
To address non-monotonicity of operators, one approach
has been to employ metrics in studying
some problematic logic programs.  These ideas were taken further in
examining quasi-metrics, and
generalized ultrametrics i.e.\ ultrametrics which take values
in an arbitrary partially ordered set (not just in the non-negative
reals).  Seda and Hitzler \cite{seda0}
``consider a natural way of endowing Scott
domains [see \cite{davey}]
with generalized ultrametrics.  This step provides a technical
tool [for finding fixpoints -- hence for analysis] of non-monotonic
operators arising out of logic programs and deductive databases and
hence to finding models for these.''

A further, similar, viewpoint is \cite{seda}:
``Once one introduces negation, which is certainly implied by
the term {\em enhanced syntax} ... then certain of the important
operators are not monotonic (and therefore not continuous), and in
consequence the Knaster-Tarski theorem [i.e.\ for fixed points; again
see \cite{davey}] is
no longer applicable to them.  Various ways have been proposed to
overcome this problem.  One such [approach is to use] syntactic conditions on
programs ... Another is to consider different operators ... The
third main solution is to introduce techniques from topology and
analysis to augment arguments based on order ... [latter include:]
methods based on metrics ... on quasi-metrics ... and finally ...
on ultrametric spaces.''

The convergence to fixed points that are based on a generalized
ultrametric system is precisely the study of
spherically complete systems and expansive automorphisms
discussed in section \ref{dilat}.  As expansive automorphisms we see here
again an example of symmetry at work.

\subsection{Link with Formal Concept Analysis}

In this subsection, we consider an ultrametric defined on the powerset
or join semilattice.  Comprehensive background on ordered sets and
lattices can be found in \cite{davey}.


As noted in section \ref{pairdiss}, typically
hierarchical clustering is based on a distance (which can be
relaxed often to a dissimilarity, not respecting the triangular inequality,
and {\em mutatis mutandis} to a similarity), defined on all pairs of the object
set: $d: I \times I \rightarrow \R^+$.  I.e., a distance is  a positive
real value.  Usually we require that a distance cannot be 0-valued unless
the objects are identical.  That is the traditional approach.

A different form of
ultrametrization is achieved from a dissimilarity defined on the power set
of attributes characterizing the observations (objects, individuals, etc.)
$X$.  Here we have: $d : X \times X \longrightarrow 2^J$, where $J$
indexes  the attribute (variables, characteristics, properties, etc.) set.

We consider a
different notion of distance, that maps pairs of objects
onto elements of a join semilattice.  The latter can represent all subsets
of the attribute set, $J$.  That is to say, it can represent the power set,
commonly denoted $2^J$, of $J$.

As an example, consider, say, $n = 5$ objects characterized by 3 boolean
(presence/absence) attributes, shown in Table \ref{tabfca}.

\begin{table}
\caption{Example dataset: 5 objects, 3 boolean attributes.}
\begin{center}
\begin{tabular}{cccc}
   &  $v_1$  &   $v_2$  & $v_3$  \\
a  &    1    &    0     &   1    \\
b  &    0    &    1     &   1    \\
c  &    1    &    0     &   1    \\
e  &    1    &    0     &   0    \\
f  &    0    &    0     &   1    \\
\end{tabular}
\end{center}
\label{tabfca}
\end{table}

Define dissimilarity between a pair of objects in Table \ref{tabfca} as
a {\em set} of 3 components, corresponding to the 3 attributes, such that
if both components are 0, we have 1; if either component is 1 and the
other 0, we have 1; and if both components are 1 we get 0.  This is the
simple matching coefficient \cite{jan1}.  We could
use, e.g., Euclidean distance for each of the values sought; but we prefer
to treat 0 values in both components as signaling  a 0 contribution.  We get
then:
\medskip
$d(a,b) = 1, 1, 0$

$d(a,c) = 0, 1, 0$

$d(a,e) = 0, 1, 1$

$d(a,f) = 1, 1, 0$

$d(b,c) = 1, 1, 0$

$d(b,e) = 1, 1, 1$

$d(b,f) = 1, 1, 0$

$d(c,e) = 0, 1, 1$

$d(c,f) = 1, 1, 0$

$d(e,f) = 1, 1, 1$

\medskip

If we take the three components in this distance as $d1, d2, d3$,
and considering a lattice representation with linkages between
all ordered subsets where the subsets are to be found in our results above
(e.g., $d(c,f) = 1, 1, 0$ implies that we have a lattice node
associated with the subset $d1,d2$), and
finally such that the order is defined on subset cardinality, then we see
that the representation shown in Figure \ref{figfca} suffices.

\begin{figure}
\begin{verbatim}                                                                                
                                                                                                
                                                                                                
Potential lattice vertices      Lattice vertices found       Level                              
                                                                                                
       d1,d2,d3                        d1,d2,d3                3                                
                                         /  \                                                   
                                        /    \                                                  
  d1,d2   d2,d3   d1,d3            d1,d2     d2,d3             2                                
                                        \    /                                                  
                                         \  /                                                   
   d1      d2      d3                     d2                   1                                
                                                                                                
\end{verbatim}

   The set d1,d2,d3 corresponds to:     $d(b,e)$ and $d(e,f)$

   The subset d1,d2 corresponds to:     $d(a,b), d(a,f), d(b,c),
                                        d(b,f),$ and $d(c,f)$

   The subset d2,d3 corresponds to:     $d(a,e)$ and $d(c,e)$

   The subset d2 corresponds to:         $d(a,c)$

\medskip

   Clusters defined by all pairwise linkage at level $\leq  2$:

$   a, b, c, f$

$   a, e$

$   c, e$

\medskip

   Clusters defined by all pairwise linkage at level $\leq 3$:

$   a, b, c, e, f$

\caption{Lattice and its interpretation, corresponding to the data shown in
Table \ref{tabfca} with the simple matching coefficient used.  (See text for
details.)}
\label{figfca}
\end{figure}

In Formal Concept Analysis \cite{davey,ganter,janowitz},
it is the lattice itself which is
of primary interest.  In \cite{jan1} there is discussion of, and a range
of examples on,  the close
relationship between the traditional hierarchical cluster analysis
based on $d: I \times I \rightarrow \R^+$, and hierarchical cluster
analysis ``based on abstract posets'' (a poset is a partially ordered
set), based on $d: I \times I \rightarrow 2^J$.  The latter, leading to
clustering based on dissimilarities, was developed initially
 in \cite{jan0}.

\section{Conclusion}

Data analysis allows us to go from measured data to a computational
path or a set of approximations used to represent the objects of
analysis.  We have noted that examples of application to face
recognition and to documents can be seen in \cite{murtult}. 

Computational logic in an analogous way used metric and ultrametric 
embeddings.  Within such topologies, computation is carried out.  
We have focused in this article on ultrametric embedding, i.e.\
given as a hierarchy or tree. 

It is interesting, and without question exciting, to envisage further 
cross-linkage between data analysis and computational logic.  

\bibliographystyle{plain}
\bibliography{symmetry-logic}

\end{document}